\documentclass[a4paper]{mem}
\usepackage{natbib}
\usepackage{graphicx}
\usepackage[a4paper]{hyperref}
\idline{73}{23}
\begin{document}
   \title{The Large Synoptic Survey Telescope (LSST) and its Impact on Variable Star Research}

   \author{A.R. Walker
}

   \offprints{A.R. Walker}
\mail{Casilla 603, La Serena, Chile}

   \institute{Cerro Tololo Inter-American Observatory, National Optical Astronomy
Observatory, Casilla 603, La Serena, Chile. 
\email{awalker@ctio.noao.edu}\\}

   \abstract{The Large Synoptic Survey Telescope ( LSST) 
   is conceived as an 8.4-m telescope with CCD or CMOS focal plane covering most of a  field
   0.6 m in diameter, the latter exceeding the size of the largest
   photographic plates ever used in astronomy.  The telescope design is driven by the  desire
   to cover much of the accessible sky to faint magnitudes on a timescale of $\sim$ one night,
   with major science drivers the search for near Earth-orbit asteroids down to a diameter
   of 300 m, discovering supernovae to z $\sim 2$, and mass tomography of the Universe via
   gravitational weak lensing.  Here the suitability of this facility for more ``traditional'' variable
   star research is investigated.

}
   \authorrunning{A.R. Walker}
   \titlerunning{The LSST and Variable Stars}
   \maketitle
%

\section{Introduction}

The origin of the LSST concept evolved from discussions on quantifying the dark matter
content of the Universe by mapping the gravitational lensing signal from faint galaxies over
very wide areas of the sky.  At this stage, the telescope concept was refered to as the
Dark Matter Telescope (DMT).  It was soon realized that two other science projects of
high interest required much the same kind of facility, these projects are the search for
Near Earth Orbiting asteroids (NEO's) down to a diameter of a few hundred meters, and
derivation of the equation of state of the Universe, so-called mapping the dark energy 
of the Universe, accomplished using type Ia supernovae as standard candles, over 
a wide range of red-shift.  These projects serve to define a telescope concept, a tentative 
experiment design, and allow estimates of the data rates and information  database 
requirements.  As is the case for the Sloan Digital Sky Survey (SDSS), telescope construction
and operation, the instrument, and data handling are tasks of similar complexity and
cost.
All these aspects of the project will be refined over the next two years 
under the guidance of the LSST Science Working Group, chaired by Michael Strauss 
(Princeton), with a view to having a detailed proposal ready for the funding agencies
 by 2004.  This early part of the project is being funded by the National Optical
Astronomy Observatory (NOAO).  A top-level rationale for LSST can be found
in the influential report ``Astronomy and Astrophysics in the New Millennium'',
see  URL http://www.nap.edu/books/, while
much information, and links to conference papers and technical reports, can be 
found at URL's http://www.noao.edu/lsst/ and http://www.lsst.org/.  Many of the
numbers, and the operations concepts for the LSST that are described below, originate from
these sources.

\section{The Concept}

The LSST telescope optical concept is based on the \citet{paul} design, which uses 
three mirrors to produce a large, well-corrected field.  An important development 
was that of \citet{willstrop}, who describes
a three-mirror telescope with perforated primary, which allows a very compact 
configuration.  He actually 
built such an instrument, which can be viewed in the grounds of the Institute of Astronomy,
University of Cambridge (U.K).
The LSST optical design is described by \citet{angel}, with important modifications by
\citet{seppala}, this has excellent
image quality  over a field of diameter three degrees;  
thin refractive elements remove residual aberrations, and the most recent design
has a flat field, which simplifies construction of the detector mosaic.
The LSST primary mirror is 8.4 m diameter,
 this being the largest size borosilicate mirror that can be produced by the University
 of Arizona Mirror Laboratory.  

To cover the large focal plane and provide well-sampled images requires of order 48Kx48K 15
micron pixels,  this could for example be achieved with 
a mosaic of 568 2K CCDs. A quick calculation suggests that this number
may not be too different than the total number of CCDs in service at professional observatories 
world-wide!  Such a large mosaic, along 
with the desire for a read-out time of a few seconds, implies a massively parallel readout
structure, and use of ASIC's, pioneered by Rockwell Scientific Co. for IR arrays, looks to be an elegant 
prospect for this purpose.  More conventional highly parallel controllers, such as the 
MONSOON controller under development by NOAO, are also a possibility.

To reach the science goal of being able to image a large fraction of the accessible sky 
to faint magnitudes in 
one night requires short exposures, even with such a large field, and $20-25$ second exposure 
times are contemplated.  Assuming a five second read time, with telescope moving to
its next target
within this period, allows the data volume and rate to be calculated: 
each read is 5 GB, so in one night 6 TB could be generated, and 1-2 PB per year.  
Processing the data in real time requires the pipeline throughput to exceed 200 MB/s.  
Although by observational astronomy standards these numbers are high, they are comparable with 
those handled by theoreticians modeling with large three-dimensional grids on super-computers.
The high pipeline speed and near-real time availability requirements demands
a system of high integrity and redundancy, 
and although the extrapolated increase of computer power, data storage and network speeds
will provide the requisite raw performance, providing usable data reliably is likely the
greatest challenge the LSST project faces.

\section{Comparison with existing Wide Field Imagers on large telescopes}

Most of the present generation of 6-10 m class telescopes do not offer a wide-field imaging 
capability, Subaru is an exception with the 10Kx8K Suprime-Cam operating at prime 
focus.  Magellan IMACS, an imaging spectrograph now being commissioned, will also provide
a relatively wide-field, 
while the Large Binocular Telescope (LBT) will offer two prime-focus cameras, 
one  optimized for the red and the other for the  blue wavelength range.  In
all cases the imagers cover approximately 0.25 sq. deg., and since they compete for telescope
time with other instruments, they are best suited for low duty-cycle programs.

As the 6-10 m class telescopes, each with several state-of-the-art instruments, take 
the place of the previous generation of 3-5 m telescopes, the tendency is for the smaller
telescopes
to become more specialized, generally also consistent with the not uncommon desire to lower 
operating costs for these older facilities.  Their operational mode ranges from classically or service 
scheduled with many runs of a few nights each, to a combination of large multi-year 
projects and smaller ``principal investigator'' projects.  Examples of the former are the SuperMACHO 
and the W Projects being
scheduled over five years at the CTIO Blanco 4-m telescope, additional
examples are  most of the other Survey projects
that are scheduled on NOAO telescopes, see http://www.noao.edu/gateway/surveys/.  
The present generation of 
CCD imagers have typically 8Kx8K 
pixels (8Kx12K on CFHT) but with 16x16K imagers either being built or planned for some
of the facilities.  

In a parallel development, only mentioned briefly here, infrared arrays have grown to
2K x 2K and have already appeared in wide-field instruments 
(e.g. U. Florida Flamingos, 
CTIO ISPI, Cornell WIRC) and are being scheduled to undertake a similar mixture of large and 
small science projects in much the same way as are the big CCD imagers.  Still larger 
instruments (eg NOAO's NEWFIRM
 4Kx4K, VISTA Project 8Kx8K) are underway, building up large mosaics using 2Kx2K 
building-blocks.

Facilities designed for ``survey science'', of which LSST will be one,
are really
experiments rather than general user facilities; data-mining 
of the archived results will be the main mode of use, in addition to the principal experiments. 
The concept of a large and expensive facility as an experiment is one that is still somewhat
alien to ground-based astronomical thinking, but very familar to the space community,
and to particle physicists.
2MASS and SDSS are present-day LSST precursors in terms of their scope and aims.
OGLE, designed originally as a gravitational lensing experiment, and now as OGLE-III
using a dedicated 1.3-m telescope with 8Kx8K CCD imager sited at Las Campanas, is
the benchmark by which the use of ``survey science'' facilities for variable star research
should be rated: 268,000
candidate variable stars in the Galactic Bulge and the Magellanic Clouds (MC), see 
e.g. \citet {wozniak}.
As pointed out by \citet {djorgovski}, surveys, by dint of their exploration of new 
parameter space sometime uncover the unexpected; rare variables such as the apparently 
normal main sequence F stars that flare by several magnitudes is one example he gives.  

With this brief summary of existing and soon-to-be-built facilities, and neglecting
some experiments in planning that may scientifically compete (PanStars, SNAP),
we now turn to address the question:
How will the LSST impact variable star research?

\section{LSST and variable stars}  

Variable stars have been traditionally broadly divided \citet{kukarkin} between
eruptive variables, periodic variables, and eclipsing variables, and are useful for (inter alia) 
(i) the determination of fundamental characteristics such as masses, radii, temperatures;
(ii) the comparison and
calibration of evolutionary models; (iii) determining the internal structure of stars; (iv) 
characterizing stellar populations; (v) distance indicators.
For eruptive variables, such as the SN science that is one of the main purposes of the telescope, LSST 
will of course efficiently discover them, with follow-up mostly or totally accomplished by other facilities, and 
one can easily imagine a greatly increased demand for follow-up facilities, especially 
those with a spectroscopic capability.  With the advent on the (Inter-)National Virtual 
Observatory (NVO) the process of finding all known information about a particular object 
will become far more efficient than at present. Clearly, near real-time availability
of both the LSST ``trigger'' and the archive information will be essential to allow
efficient follow-up.  For study of non-eruptive variables, the utility of LSST is not
so clear, and it is first necessary to summarize the envisaged mode of operation.

The strawman LSST operations plan is to operate in two survey modes: (i) An Ultra-Deep 
multi-band probe mode, covering 1000 square degrees to 29th magnitude, with 10 hours 
exposures.  This would determine weak lens tomographic mass versus cosmic time, and 
also discover 3000 high-z supernovae per year, and identify faint transients; and (ii) an 
All-Sky Synoptic Survey Mode, reaching 24th magnitude at 10-sigma with 20 second 
exposures, this survey would identify NEO's with diameter greater than 300 m, 
200,000 local supernovae, and also identify faint transients.  Preceding the start
of the two surveys, a full year would be spent covering 14,000 square degrees to   
26th magnitude in g,r,i,z, to provide a reference.  In subsequent years 25 percent of the time will 
be spent on the Ultra-Deep survey, and 75 percent of the time on the All-Sky survey, the latter using only the 
r-filter.  It is instructive to calculate the efficiency of the All-Sky survey.  Using 
figures representative for Cerro Tololo and Cerro Pachon, we assume operation for
200 dark-grey hours per month, 75 percent photometric or usable, and of these nights
 85 percent have useable seeing, and of this the telescope uptime is
95 percent. Thus with 75 percent for the All-Sky survey, 
this gives 91 hours/month, and with 30 seconds between exposures (exposure time plus all overheads) 
the sky coverage is 63,300 square degrees/lunation.  If the whole available sky (14,800 square degrees 
nominally) is to be covered this corresponds to only four visits, so there is an immediate trade to be 
made between area coverage and re-visit rate.  Although the desirability of putting the telescope on a highly 
photometric site cannot be disputed, there are many open questions remaining to be 
answered concerning techniques and strategy.  For instance, it is not envisaged to use the telescope
for these surveys near full Moon.  Disregarding possible difficulties in baffling the telescope
against stray light, the difference in S/N between new Moon and full Moon through an r filter 
of a 24-th magnitude star in a 20 second exposure, 
assuming 0.9 arc sec seeing,
is only a reduction from S/N 10 to 7; the critical factor 
with a brighter sky is the seeing.  So although this restriction may be replaced by a more
complicated one involving a combination of sky brightness and seeing, we will return below to utilization
of this possible niche.
Clearly the diverse main science projects 
make optimization a difficult task, without even considering any constraints that
might be desirable in order to allow other
science projects.

We are now in a position to make a general evaluation of the usefulness of the LSST for periodic 
variable star studies.
For objects on the sky less than a few square degrees in area (i.e. up to the MC in size) 
targeted surveys such as OGLE and Super-MACHO will be more efficient at the largest size, 
with a whole range of facilities able to better study smaller objects such as Local Group
dwarf galaxies, see 3. above.  Where the LSST will be supreme is in its ability
to rapidly survey huge volumes of our galaxy, for instance it
will allow study of the galactic halo structure via RR Lyraes, and even with
the short exposures envisaged will have 
the sensitivity to discover these stars out to $\sim 500$ Kpc.  Any halo structure that is 
old and thus containing RR Lyraes will be able to be measured and mapped.  It will efficiently
survey the galactic disk, although how far it pushes to low galactic latitudes has not 
been decided.  It will be able to discover, and follow quite efficiently, long-period
variables such as Miras.  It will find myriads of cataclysmic variables (CV's), and should be able
to overcome the biases inherent in surveys for these objects to date, allowing an objective
comparison with CV formation theory, at present very discordant with the observations.  
As has been stressed above, having
facilities available for follow-up, both photometric and spectroscopic will be essential,
together with the availability of
multi-wavelength cross-identifications to filter out non-interesting objects.  Much of what
is said above is also generally applicable to eclipsing binaries, presently of much interest
after the discovery of sub-stellar companions  by \citet{henry} via this technique.

It should be stressed that the usefulness of the LSST for traditional variable star studies
is restricted by the operations model, not the telescope design itself.  For instance,
the LSST can cover 28 sq. deg., i.e. most of the LMC, in 4 exposures, and with 0.9
arcsec seeing, S/N $\sim 25$ is achieved in B and R at magnitude 23 in 60 seconds, even at full Moon;
the LMC could be surveyed in two filters to this magnitude in under 10 minutes!
Thus the operational niche that should be explored is that where the sky is bright and
perhaps the seeing in the worst quartile.  This parameter space will not be without
competition, as I and z band photometry will be even less dependent on lunar phase than
shorter wavelength bandpasses.

\section{Conclusions}

The $2-8$ m class telescopes plus $8-16$K imagers will continue to do the ``few square degree'' 
science, operating in both conventional few hour - few night runs, and in longer ``survey'' modes.
Pure survey-mode facilities also do some of this science, but as they tackle projects requiring
more and more area, the repeat rate per field  drops rapidly towards one.  The LSST will
open new 
parameter-space - deep and wide and fast, and it is likely to
discover new classes of variables, where follow-up by other facilities will
be essential.
Its periodic-variable targets are those in the Galaxy, 
since even for structures several degrees across, such as the MC, other facilities performing
targeted studies will do better, by trading field for time. 

The LSST Project is in definition phase, and now is the time for scientific input, via
the LSST 
Science Working Group, which will define the science cases which will flowdown into 
the technical requirements.  Questions as to the optimum number of filters, the required 
calibration accuracy, the photometric depth needed - all are awaiting detailed
treatment. While the thrust for the main scientific projects of the LSST should not be
diluted, some parameters of the design may be able to be tweaked to enable other science.
In particular, use of the time around full Moon, and poorer seeing conditions, are a
parameter space that could be exploited for variable star programs of many types.

\bibliographystyle{aa}

\end{document}